\documentclass[reprint,amsmath,amssymb,nofootinbib]{revtex4}
\usepackage{ulem}
\usepackage{amssymb}
\usepackage{amsmath}
\usepackage{graphicx}
\usepackage{dcolumn}
\usepackage[colorlinks,urlcolor=blue,citecolor=blue,linkcolor=blue]{hyperref}
\usepackage{color,units}
\usepackage[dvipsnames]{xcolor} 
\usepackage{lineno}
\usepackage{xspace}
\usepackage{longtable} 
\usepackage{float} 
\usepackage{multirow}
\usepackage{amsfonts,wasysym,epsfig,verbatim,subfigure,bm,mathrsfs,lipsum}

\begin{document}

\newcommand{\GSSI}{Gran Sasso Science Institute (GSSI), I-67100 L’Aquila, Italy}
\newcommand{\GranSasso}{INFN, Laboratori Nazionali del Gran Sasso, I-67100 Assergi, Italy}

\newcommand{\MPI}{Max-Planck-Institut f{\"u}r Gravitationsphysik (Albert-Einstein-Institut), D-30167 Hannover, Germany}

\newcommand{\LBNZ}{Leibniz Universit{\"a}t Hannover, D-30167 Hannover, Germany}

\title{Eccentricity evolution of spinning binaries and its dependence on the equation of state of the components}
\author{Sayak Datta}\email{sayak.datta@gssi.it}
\affiliation{\GSSI}\affiliation{\GranSasso}\affiliation{\MPI}\affiliation{\LBNZ}
\date{\today}

\begin{abstract}
    We study the evolution of the eccentricity of an eccentric orbit with spinning components. We develop a prescription to express the evolving eccentricity in terms of reference eccentricity and frequency. For that purpose we considered the spins to be perpendicular to the orbital plane. Using this we found an analytical result for the contribution of spin in eccentricity evolution. As a result, we expressed orbital eccentricity in a series of reference eccentricity and gravitational wave frequency. The prescription developed here can easily be used to find arbitrarily higher-order contributions of reference eccentricity. With this we computed the eccentricity upto $\mathcal{O}(e_0^5)$. This result can be used to construct the waveforms of spinning compact objects in an eccentric orbit. Since, our expression depends on the spin induced quadrupole moments, we also study the impact of component properties on the eccentricity evolution through the quadrupole moment. We find for BNSs the evolution depends on the equation of state very mildly unless the NSs are subsolar mass. For subsolar mass NSs the deviations from BH case is comparatively larger and has equation of state dependence. For binary boson stars the deviations are comparatively larger across the mass values. We argue that it may affect our understanding of formation channels and their corresponding populations. We also argue that this can possibly be used as another tool to constrain exoticness of compact objects in a binary.
\end{abstract}

\maketitle

\section{Introduction}

In recent times, the detection of gravitational waves (GWs) from the coalescence of compact binaries with ground-based detectors \cite{TheLIGOScientific:2014jea, TheVirgo:2014hva} has opened up a new era of astronomy~\cite{LIGOScientific:2018mvr, LIGOScientific:2020ibl}. Most of the sources are believed to be black hole (BH) binaries. The merger of neutron stars (NSs) was also observed in the event GW170817~\cite{gw170817}, and possibly also GW190425~\cite{Abbott:2020uma}. More recently, detections of GW200105 and GW200115~\cite{bhns_LIGOScientific:2021qlt} were also made where it is believed that it is BH-NS binary. Currently, the existing detectors are continuously being upgraded. Alongside, there are proposals for several next-generation ground-based detectors such as the Einstein telescope \cite{maggiore2020science} and cosmic explorer \cite{Reitze:2019iox}. These detectors will be significantly more sensitive compared to the current detectors. As a result, it will be possible to measure very small features in the signals. Similarly, space-based detectors such as Laser Interferometer Space Antenna (LISA) \cite{LISA:2017pwj} are also being built. LISA will observe binaries comprising supermassive bodies. These sources will be either very loud or will last very long in the detector for the detector to measure very small features in the signal. Therefore modeling the signals as accurately as possible has become a necessity.

In the context of GW astronomy, primarily the focus has been on the circular orbits. This is reasonable as we expect the stellar mass binaries to have low eccentricities \cite{Peters:1964zz, Peters:1963ux}. However, compact binaries that formed via the dynamical interactions in dense stellar environments or through the Kozai-Lidov processes \cite{Kozai:1962zz, lidov1962evolution}, are expected to retain residual eccentricities $e_0\gtrsim .1$ during observation with ground-based detectors \cite{LIGOScientific:2019dag}. Therefore, the measurement of nonzero eccentricities of the binaries may shed light on our understanding of the formation channels. On the other hand extreme mass ratio inspirals are also expected to have large eccentricities in the observable band \cite{Amaro-Seoane:2012lgq}. In general, binaries can have environment around it as well.  As a consequence, the modified geodesic motions, environmental forces such as dynamical friction and accretion can lead to significant eccentricity growth (see \cite{Cardoso:2020iji} and references therein). The modification of the eccentricity evolution can also arise due to the existence of dark matter and beyond standard model physics. This effect has been observed in binaries in the presence of scalar environments \cite{Tomaselli:2024bdd, Tomaselli:2024dbw, Boskovic:2024fga}. Hence, it is crucial to understand and model the effects of eccentricity on a binary.

Interestingly, some of the recent analyses \cite{LIGOScientific:2020ufj, Kimball:2020qyd, Romero-Shaw:2021ual, Romero-Shaw:2022xko, OShea:2021ugg}
support the presence of eccentricity in the observed binary black hole (BBH) events. Alongside, it was argued in Ref. \cite{Favata:2021vhw}, that the presence of even smaller eccentricities $e_0 \sim .01-.05$ may induce systematic biases in parameter estimation analyses. Since the next generation ground-based detectors will have improved low-frequency sensitivity compared to the current generation detectors, these detectors can make confident observations of eccentric systems \cite{Lower:2018seu}.

There have been efforts in the past to model the inspiral waveforms from compact binary mergers \cite{Mishra:2015bqa, Moore:2016qxz, Tanay:2016zog, Boetzel:2019nfw, Ebersold:2019kdc, Konigsdorffer:2006zt, Moore:2019xkm, Paul:2022xfy, Datta:2023wsn}. Although the effect of spins has been modeled within the post-Newtonian formalism \cite{Marsat:2012fn, Hartung:2011te, Levi:2015msa, Bohe:2012mr, Bohe:2013cla, Marsat:2013caa, Bohe:2015ana, Blanchet:2011zv, Chatziioannou:2013dza, Henry:2022dzx, Arun:2008kb}, a combined treatment including spins and eccentricity is largely absent.
There have been efforts such as those of Ref. \cite{Kidder:1995zr, Buonanno:2012rv, Majar:2008zz, Klein:2010ti, Klein:2018ybm, Klein:2021jtd, Blanchet:2008je, Kidder:2007rt, Blanchet:2013haa, Phukon:2019gfh} that attempt to address this concern to some extent. However, more progress is needed.

Although there are some works on eccentric waveforms, mostly they do not consider the effect of the interaction between spin and eccentricity. Keeping this in mind, we will try to find the eccentricity evolution. In Ref. \cite{Klein:2010ti} the equations governing the time evolution of the orbital parameters, and in particular the eccentricity, including the spin-orbit and the spin-spin couplings were computed. These are needed to accurately compute the post-Newtonian approximation up to 2PN accuracy. However, these results are not sufficient to compute GW waveforms. The computed results depend on the ``instantaneous" eccentricity of the orbit. However, in the inspiral time scale, the eccentricity of an orbit evolves. Hence, it is important to express the instantaneous eccentricity in terms of some ``reference eccentricity" and GW frequency. 
In the leading order of reference eccentricity for nonspinning orbit this was explored in a few works \cite{Arun:2009mc, Moore:2016qxz,Moore:2019xkm}. Some, higher order terms of reference eccentricity were computed in ref. \cite{Yunes:2009yz}, albeit for non-spinning binaries. However, in ref. \cite{Datta:2023wsn} a more comprehensive prescription was developed for both nonspinning and spinning binaries, that can be used to iteratively find the arbitrarily higher order terms in reference eccentricities and GW frequency.

In the current work we extend the results of Ref. \cite{Datta:2023wsn} to spinning binaries with non-zero quadrupole moment (QM), to higher PN order and also different parametrization $(y)$ defined below. Therefore, it will be for the first time that the eccentricity evolution of spinning binaries will be analytically expressed in terms of reference eccentricity and $y$. This as a result provides eccentricity-spin coupling terms in fluxes. We develop the prescription for spins perpendicular to the orbital plane. The prescription discussed here can be extended to arbitrarily high order in eccentricity, in principle. Therefore these results can be directly used to model the GW waveforms for spinning bodies in an eccentric orbit. With the newly found expressions, we also study the impact of the component's equation of state (EoS) on the eccentricity evolution. This happens due to the eccentricity-quadrupole moment (E-Q) coupling in the evolution. This can have impact on our understanding of formation channels and also can pave new ways to test the nature of the compact objects.

\section{Equations of eccentric orbits}
\label{sec:eom}

Our purpose in the current work is to formulate a method that can be used to analytically compute the eccentricity of an orbit with spinning components with respect to the emitted GW frequency. As a result, this can be used to compute frequency-dependent fluxes and consequently the GW waveform. For this purpose we will primarily follow the notations and discussions in Ref. \cite{Klein:2010ti, Klein:2018ybm}. The spin-orbit couplings appear at 1.5PN order and spin-spin couplings appear at 2PN order. For this reason Ref. \cite{Klein:2010ti, Klein:2018ybm} considered only the Newtonian and spin-coupling terms in the equations of motion. For simplicity, we will use a system of units where $G=c=M=1$, where $M$ is the total mass of the system. The boldface represents a $3-$vector and a hat above represents a unit vector. The Lagrangian of the system is provided in Ref. \cite{Klein:2018ybm, Kidder:1992fr, Kidder:1995zr}.

A quasi-Keplerian solution to the equation of motion derived from the Lagrangian is described in Ref. \cite{Klein:2018ybm}.

\begin{eqnarray}
    r =& a(1-e_r \cos{u}) + f_r(v))\\
    \phi =& (1+k)v + f_{\phi}(v) \\
    v =& 2\arctan \Big(\sqrt{\frac{1+e_{\phi}}{1-e_{\phi}}}\tan \frac{u}{2}\Big)\\
    l =& u -e_t \sin{u} + f_t(u,v)\\
    \Dot{l} =& n,
    \label{eq:orbital-quantities}
\end{eqnarray}

where $(r,\phi)$ is a polar coordinate system in the plane of motion, $n$ is the mean motion, $u$, $v$, and $l$ are the eccentric, true, and mean anomalies, a is the semi-major axis, $e_t$, $e_r$, and $e_{\phi}$ are eccentricities, $k$ is the perihelion precession, and the $f_i$ are constants \cite{Klein:2010ti}.

Following \cite{Klein:2018ybm}, we will express our equations in terms of the post-Newtonian (PN) parameter $y$ defined as,

\begin{equation}
    y= \frac{[M(1+k)n]^{1/3}}{\sqrt{1-e_t^2}}
\end{equation}
We will treat $y$ as a perturbative variable that keeps track of the PN order. At the innermost stable circular orbit (ISCO) it takes the value $y_{ISCO}=1/\sqrt{6}$. For circular orbits in the leading order $y^3 \sim Mn \sim v^3$, where $v$ is the post-Newtonian velocity parameter. The spin and eccentricity dependence of several orbital quantities were found in Ref. \cite{Klein:2018ybm}. We only show,

\begin{widetext}
    \begin{align}
    a (1-e_t^2)y^2=&  1+ \left(-1+\frac{\nu}{3} + (3-\frac{\nu}{3})e_t^2\right)y^2 +\beta\left(\frac{2}{3}+2e_t^2, 1+ e_t^2\right)y^3 +\Bigg[5+\frac{11}{4}\nu +\frac{\nu^2}{9}+ \Big(\frac{21}{2}-\frac{73}{6}\nu -\frac{2}{9}\nu^2\Big)e_t^2 \\
    &+ (1+\frac{5}{12}\nu +\frac{\nu^2}{9})e_t^4 +(1-e_t^2)^{3/2}(-5+2\nu)   +\frac{\gamma_1}{2} (1+e_t^2) \Bigg]y^4  \nonumber \\
    n =& (1-e_t^2)^{3/2} y^3 \left(1-3y^2 -\beta(4,3)y^3 + \left(-\frac{9}{2} + 7\nu +(-\frac{51}{4}+\frac{13}{2}\nu)e_t^2 -\frac{3}{2}\gamma_1\right) y^4\right),
    \label{eq:a-n}
\end{align}
\end{widetext}

where,

\begin{equation}
   2 \gamma_1 = 3\left(\Hat{\bf L}.{\bf s}\right)^2- {\bf s}^2 +\sum_{i=1}^2(Q_i-1)\left[3\left(\Hat{\bf L}.{\bf s}_i\right)^2- {\bf s}_i^2 \right],
   \label{eq:gamma}
\end{equation}
with  $\hat{{\bf L}}$, ${\bf s}_i$, and  $Q_i$ are normal to the orbital plane, reduced spin of $i$th body and $i$th body's QM respectively, and ${\bf s} = {\bf s}_1 + {\bf s}_2$. 

In the quasi-Keplerian formalism, PN expansions of elliptical orbit quantities are performed most naturally in terms of the radial orbit angular frequency $\omega_r \equiv n$. This frequency is the mean motion or periastron-to-periastron angular frequency. However, the limitation of this frequency is that correspondence with the circular orbit limit is not straightforward. Unlike the eccentric case, the circular orbit quantities are more naturally expanded in terms of the azimuthal or $\phi-$angular frequency $\omega_{\phi} \equiv \xi_{\phi}/M$. Hence, $\xi_{\phi}$ is usually used for parametrizing eccentric binaries. However, following Ref. \cite{Klein:2018ybm} we use the variable $y$ defined earlier, which also serve a similar purpose.

\section{eccentricity evolution}
\label{sec:eccentricity evolution}

In this section, our focus will be on computing the eccentricity of a binary in terms of binary frequency. Currently, these results are available for non-spinning binaries \cite{Datta:2023wsn, Moore:2016qxz}. In this section, we will construct a prescription that can be used to express the eccentricity in a series summation of power of reference eccentricity and power of frequency to an arbitrary order, for a spinning binary.

\subsection{General analytical expression in a series expansion}

\begin{figure*}
\includegraphics[width=85mm]{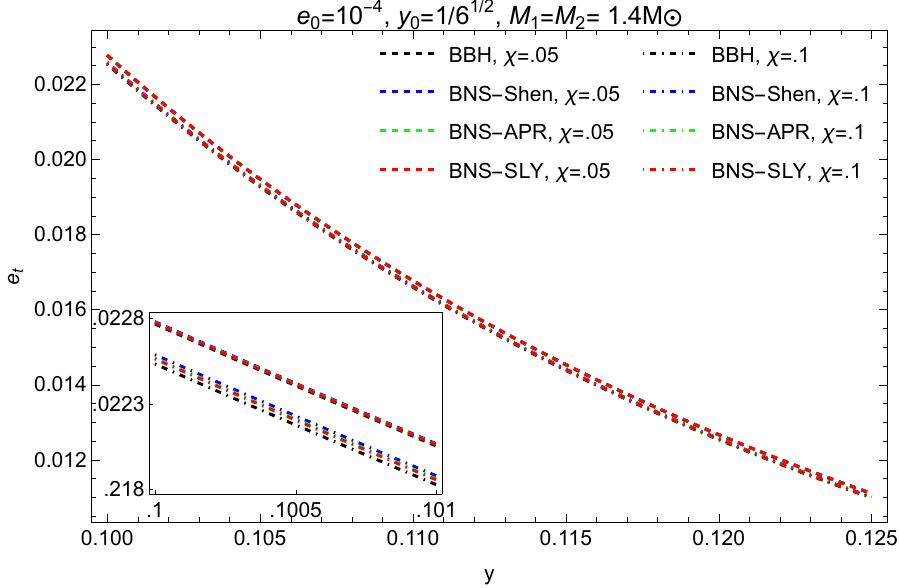}
\includegraphics[width=85mm]{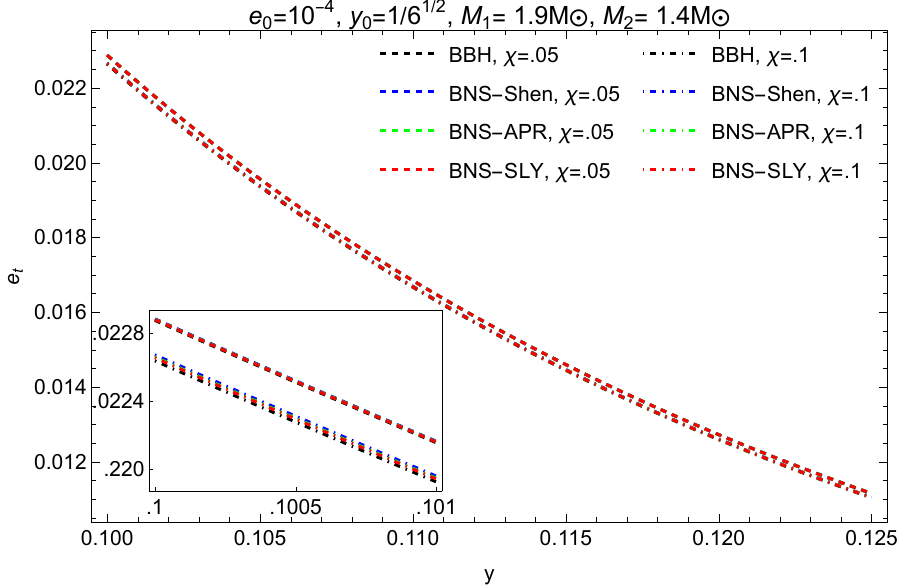}
\includegraphics[width=85mm]{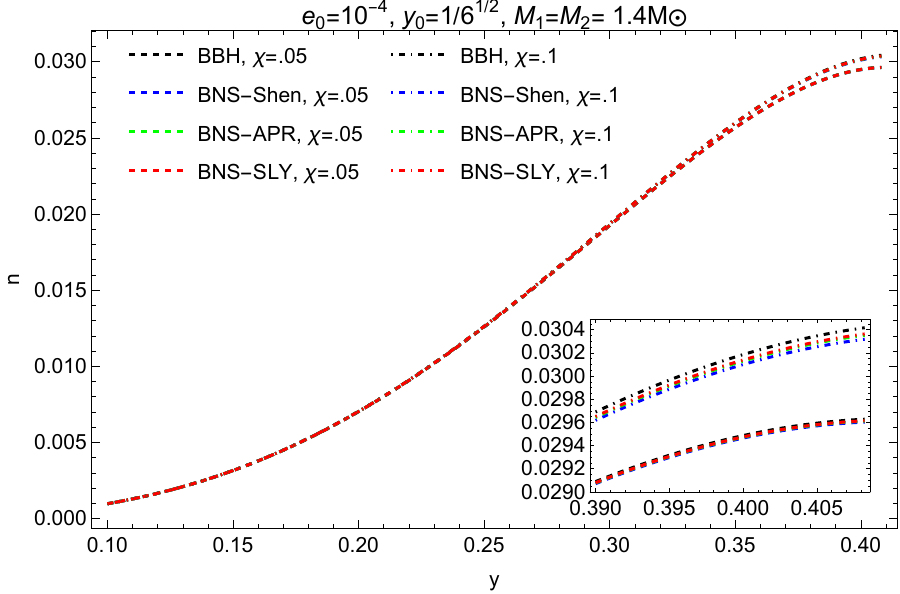}
\includegraphics[width=85mm]{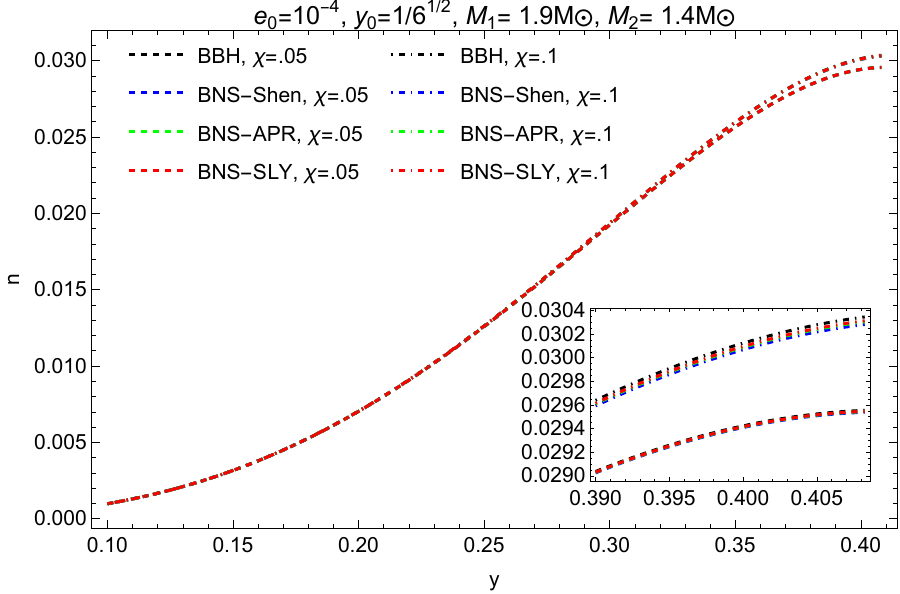}
\caption{We plot the evolving eccentricity $e_t$ as a function of the post-Newtonian parameter $y$ for BBHs (dashed and dotted black curves) and BNSs with different equations of state (red, green, and blue curves). For all systems, we set $y_0 = 1/\sqrt{6}$ at the ISCO, and evolve the binaries backward in time, i.e., toward lower $y$ values. We set the reference eccentricity as $e_0 = e_t(y_0) = 10^{-4}$. In the left panel, we consider equal-mass components with equal spins. In the right panel, we consider unequal-mass components with equal spins.
}
\label{fig:BBH-BNS}
\end{figure*}

\begin{figure*}\includegraphics[width=85mm]{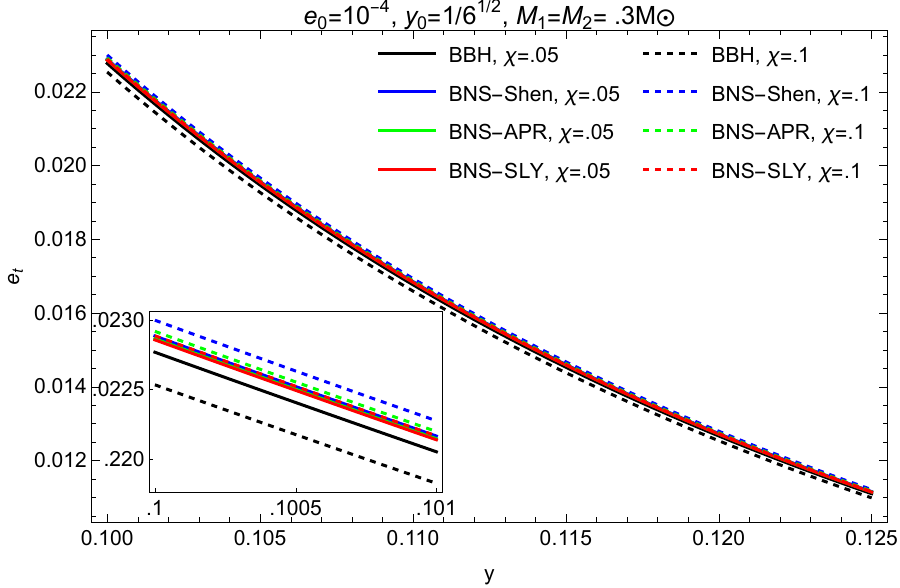}
\includegraphics[width=85mm]{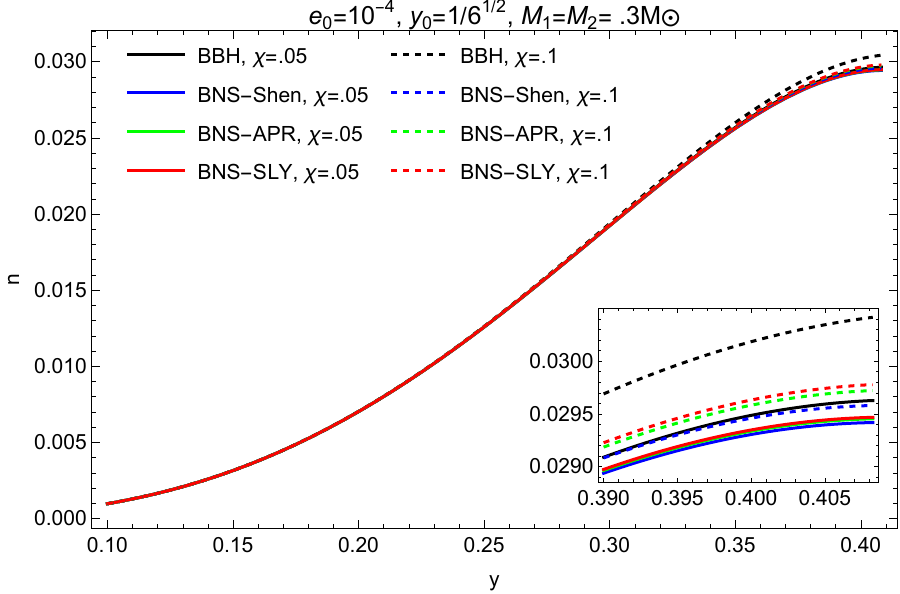}
\includegraphics[width=85mm]{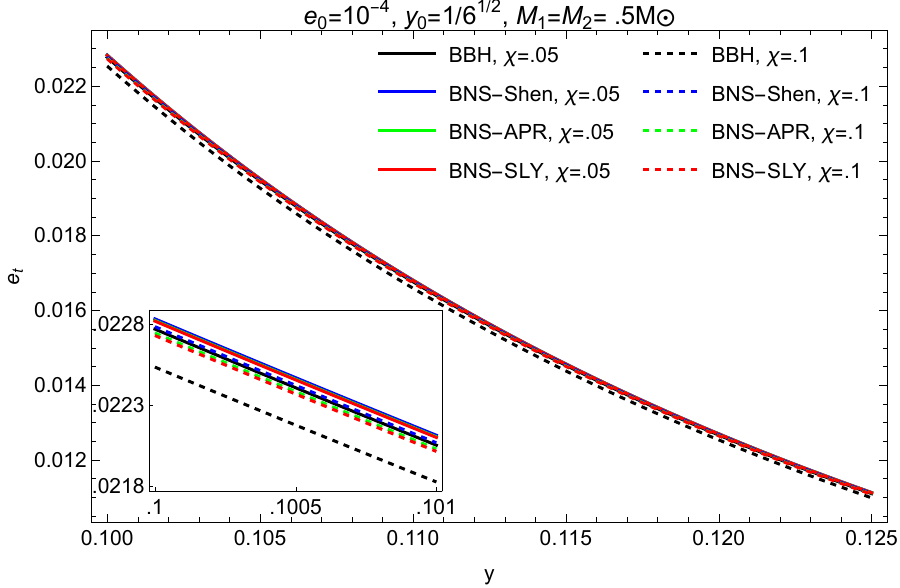}
\includegraphics[width=85mm]{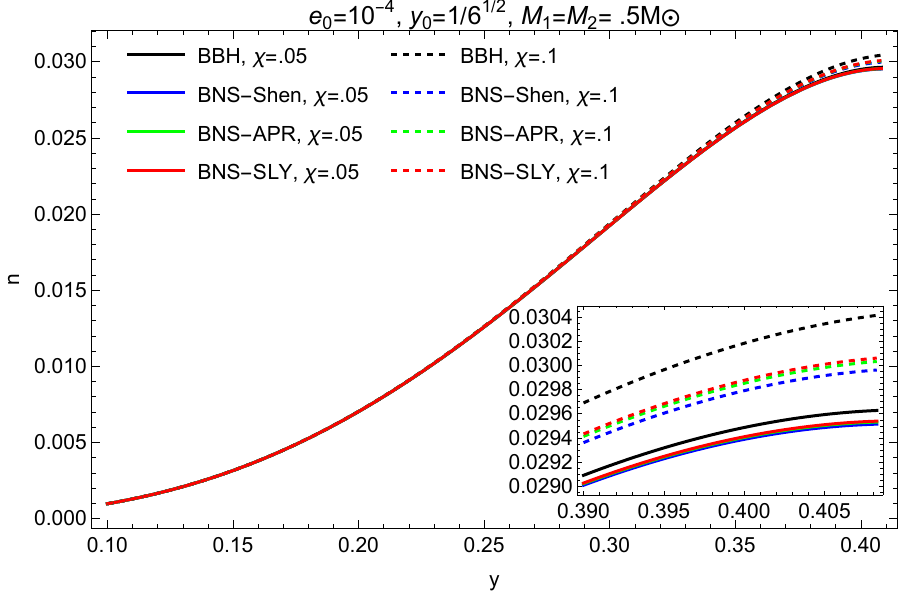}
\includegraphics[width=85mm]{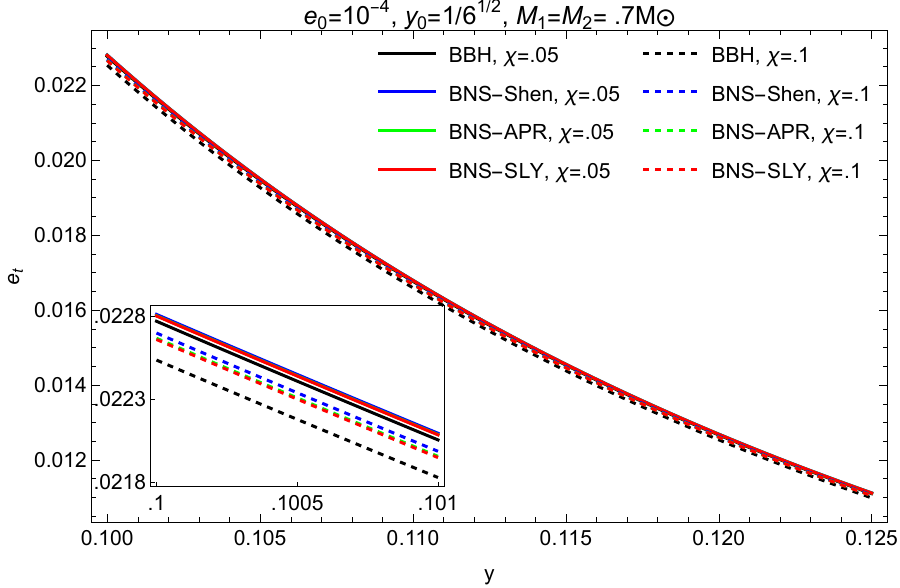}
\includegraphics[width=85mm]{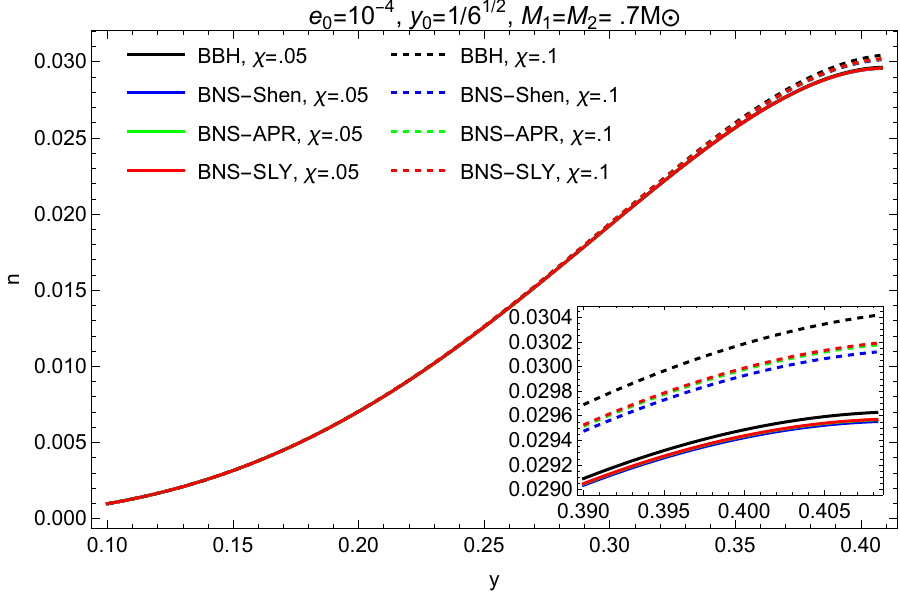}
\caption{In the above figure we consider equal mass BNS with component masses $.3M_{\odot}$, $.5M_{\odot}$, and $.7M_{\odot}$, with different equations of state (red, green, and blue curves). In the left column we show the eccentricity evolution and in the right column we show evolution of $n$. For all the cases considered $y_0 = 1/\sqrt{6}$ at the ISCO. We set the reference eccentricity as $e_0 = e_t(y_0) = 10^{-4}$. It was shown in Ref. \cite{Yagi:2013awa}, the low mass NSs has larger QM. Hence, if such low mass NSs do exist and form binaries, the EoS impact will be larger on the orbital dynamics. 
}
\label{fig:BBH-BNS-subsolar}
\end{figure*}

In Sec. \ref{sec:eom} we discussed the parametric solution in the quasi-Keplerian approach. This is done using the constants of the motions considering only the conservative part. However, during orbital motion, the system emits GWs that carry away energy and angular momentum. As a result, the conserved quantities start to change in the inspiral time scale. This corresponds to the dissipative part of the equation of motion. Conventionally it is tackled by solving the evolution equation of the conserved quantities in the inspiral time scale. For the current work, we only require the evolution equation of the $e_t$ and $y$. Their evolution equation can be expressed as \cite{Arun:2009mc, Klein:2010ti, Klein:2018ybm},

\begin{equation}
    \label{eq: e-x time derivative}
    \begin{split}
       M \frac{dy}{dt} =&  (1-e_t^2)^{3/2} \nu y^9\left(\mathcal{Y}_0 + \sum_{n=2}^6 y^n \mathcal{Y}_n \right) \\
       M \frac{de_t^2}{dt} =& - (1-e_t^2)^{3/2} \nu y^8\left(\mathcal{E}_0 + \sum_{n=2}^6 y^n \mathcal{E}_n \right),
    \end{split}
\end{equation}
where, $\mathcal{Y}_n$ and $\mathcal{E}_n$ is provided in Ref.\cite{Klein:2018ybm}, where $\mathcal{Y}_n = a_n$ and $\mathcal{E}_n = b_n$. We find $de_t/dy=(de_t/dt)/(dy/dt)$.

Once an expression of $de_t/dy$ is found, the right-hand side of the equation can be expressed in a series expansion of $e_t$ where the coefficients of each term depend on $y$. This equation then is solved to find the expression of $e_t$ in terms of an reference eccentricity $e_0$ and $y$. In this section, we will develop a prescription that can be used to find very high-order powers in $e_0$. 

To demonstrate the prescription, we will keep up to power $e_t^3$. Then the eccentricity evolution can be expressed as follows,

\begin{equation}
\label{eq:e derivative}
    \frac{de_t}{dy} =  e_t f_1(y) + e_t^3 f_3(y)+ e_t^5 f_5(y) + \mathcal{O}(e_t^6).
\end{equation}

Note, we have not explicitly specified the functional expression of $f_1(y)$ and $f_3(y)$. Once the prescription is described, we will substitute them with the required PN accuracy.

From Eq. (\ref{eq:e derivative}), the resulting solution can be found by integrating the equation on both sides. During the process, we identify that $e_t \rightarrow e_0$, i.e. the reference eccentricity, when $y \rightarrow y_0$, i.e. the reference frequency. we find,

\begin{equation}
    \begin{split}
        \int_{e_0}^{e_t}\frac{de_t}{e_t} =& \int_{y_0}^{y}dy \Big(f_1(y) + e_t^2 f_3(y) + \mathcal{O}(e_t^3)\Big)\\
    \ln(\frac{e_t}{e_0}) =&  \int_{y_0}^{y}dy f_1(y) + \int_{y_0}^{y}dy e_t^2 f_3(y) + \mathcal{O}(e_t^3)\\ 
    e_t =&  e_0e^{\int_{y_0}^{y}dy f_1(y)}  e^{\int_{y_0}^{y}dy e_t^2 f_3(y) + \mathcal{O}(e_t^3)}.
    \end{split}
\end{equation}

\begin{figure*}\includegraphics[width=85mm]{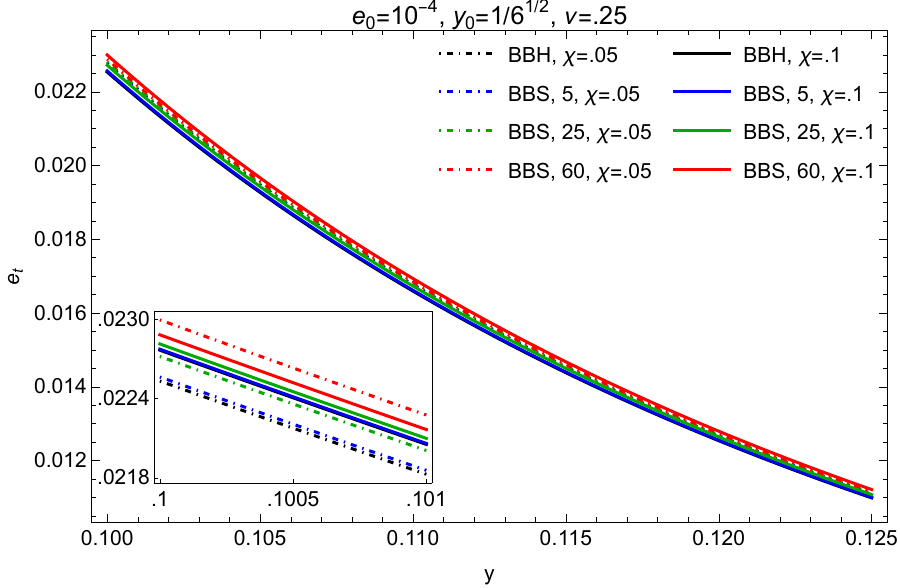}
\caption{We plot the evolving eccentricity $e_t$ as a function of the post-Newtonian parameter $y$ for BBHs (dashed and dotted black curves) and binary boson stars with different QMs. Red, green, and blue curves represent $Q=60,\, 25, \, 5$. For all systems, we set $y_0 = 1/\sqrt{6}$ at the ISCO, and evolve the binaries backward in time, i.e., toward lower $y$ values. We set the reference eccentricity as $e_0 = e_t(y_0) = 10^{-4}$. We consider equal-mass components with equal spins.}
\label{fig:BBH-BBS}
\end{figure*}

\begin{figure*}\includegraphics[width=85mm]{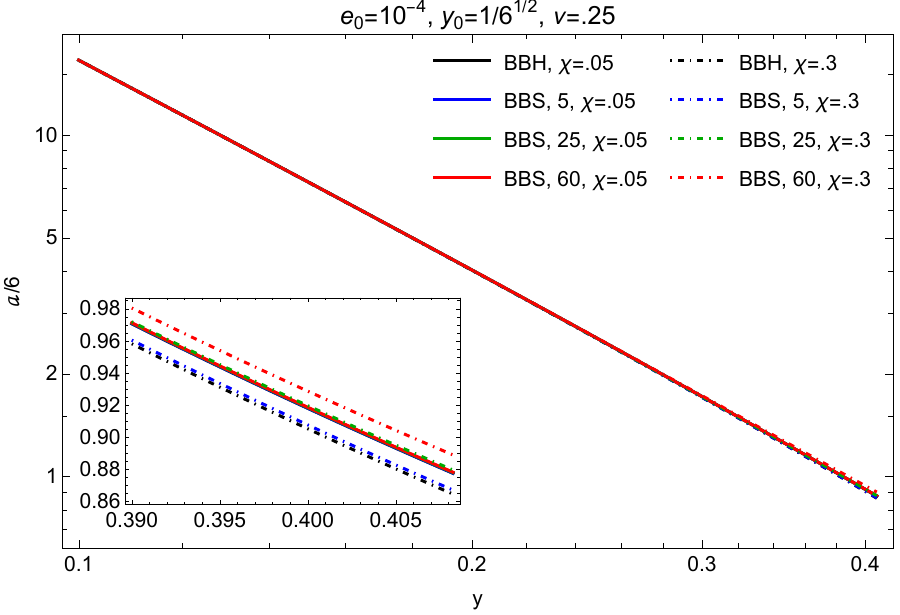}
\includegraphics[width=85mm]{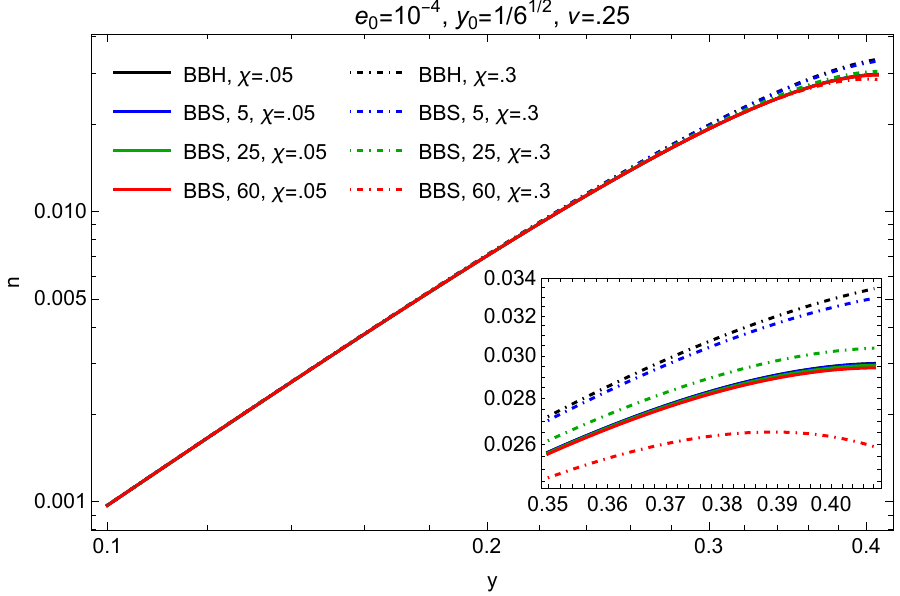}
\caption{In the above plot we show the evolution of $a$ and $n$, defined in Eq. \ref{eq:orbital-quantities}. Red, green, and blue curves represent $Q=60,\, 25, \, 5$. $a$ and $n$ deviates from BH values for larger $Q$ and $\chi$. The deviations are stronger only for larger $y$.}
\label{fig:BBH-BBS-a-n}
\end{figure*}

Note, that the first integral on the right-hand side is independent of the eccentricity while the second one depends on it. Therefore although the second integral can not be computed without the explicit knowledge of $e_t$, the first integral can be computed. Hence the expression can be rearranged as, 

\begin{equation}
    e_t =  e_0 \frac{e^{F_1{(y)}} }{e^{F_1{(y_0)}} } e^{\int_{y_0}^{y}dy e_t^2 f_3(y) + \mathcal{O}(e_t^3)}.
\end{equation}

The second integral has $e_t$ inside the integral. Therefore, without the exact knowledge of $e_t$ in terms of $y$, this integral can not be computed exactly. However, the eccentricity $e_t$ inside the second integral can be replaced with the above equation and as a result, a leading order term of the integral can be computed. Therefore, although an exact integral is not computable, an approximate result can be found which is exact to a particular order. This as a result can be used to find the next order term. This can be continued for arbitrary powers of $e_0$. In this work, we will only compute up to $e_0^5$ term as Eq. (\ref{eq:e derivative}) keeps only up to $e_t^5$. But in principle, this can be continued iteratively by considering higher order terms in Eq. (\ref{eq:e derivative}). After rearranging the expressions they can be expressed as,

\begin{widetext}
\begin{equation}
    \begin{split}
        \frac{e_t e^{F_1{(y_0)}} }{e_0 e^{F_1{(y)}}} =& e^{e_0^2\int_{y_0}^{y}dy \frac{e^{2F_1{(y)}} }{e^{2F_1{(y_0)}} } \Big(e^{2\int_{y_0}^{y}d\Bar{y} e_t^2 f_3(\Bar{y}) + \mathcal{O}(e_t^3)}\Big)\,\, f_3(y) + \mathcal{O}(e_t^3)}\\
        \frac{e_t e^{F_1{(y_0)}} }{e_0 e^{F_1{(y)}}} =& \Big(1+e_0^2\int_{y_0}^{y}dy \frac{e^{2F_1{(y)}} }{e^{2F_1{(y_0)}} } \Big(e^{2\int_{y_0}^{y}d\Bar{y} e_t^2 f_3(\Bar{y}) + \mathcal{O}(e_t^3)}\Big)\,\, f_3(y) + \mathcal{O}(e_0^3)\Big)\\
        \frac{e_t e^{F_1{(y_0)}} }{e_0 e^{F_1{(y)}}} =& \Big(1+e_0^2\int_{y_0}^{y}dy \frac{e^{2F_1{(y)}} }{e^{2F_1{(y_0)}} } \Big(1 + \mathcal{O}(e_0^2)\Big)\,\, f_3(y) + \mathcal{O}(e_0^3)\Big).
    \end{split}
\end{equation}
\end{widetext}

This as a result boils down to a series of $e_0$, where the coefficients of the expansions are integrals in $y$. This approach can be used consecutively after deriving individual coefficients of a particular order. Interestingly, to find the schematic structure of the coefficients in integral form of an arbitrary power of $e_0$ it is not required to know $f_i(y)$. The expression can be derived in terms of the integrals of $f_i(y)$s.

The above results can be expressed in a further simplified and compact form as,

\begin{equation}
\label{eq:eccentricity evolution}
        \begin{split}
            e_t =&\frac{e^{F_1{(y)}} }{e^{F_1{(y_0)}} } (e_0+e_0^3\int_{y_0}^{y}dy \frac{e^{2F_1{(y)}} }{e^{2F_1{(y_0)}} }  f_3(y) + \mathcal{O}(e_0^4)).
        \end{split}
\end{equation}

Similarly continuing this upto $\mathcal{O}(e_0^5)$ we find,
\begin{equation}
\label{eq:final et}
        \begin{split}
            e_t =& e_0\frac{\mathcal{A}_1(y)}{\mathcal{A}_1(0)} + e_0^3\frac{\mathcal{A}_1(y) }{\mathcal{A}_1(0)^3} \left[\mathcal{A}_3(y)- \mathcal{A}_3(0) \right] +e_0^5\frac{\mathcal{A}_1(y)}{\mathcal{A}_1(0)^5}[ \mathcal{A}_5 (y) - \mathcal{A}_5(0) + \frac{1}{2}\mathcal{A}_3(y)^2 -3 \mathcal{A}_3(y)\mathcal{A}_3(0)  + \frac{5}{2}\mathcal{A}_3(0)^2 ] + \mathcal{O}(e_0^7)
        \end{split}
\end{equation}

where,

\begin{eqnarray}
\label{eq:A A2 B}
    F_1(y) - F_1(y_0) =& \int_{y_0}^{y}dy   f_1(y) \nonumber\\
    \mathcal{A}_1(y) =& e^{F_1{(y)}} \nonumber
    \\
    \mathcal{A}_3(y) - \mathcal{A}_3(y_0) =& \int_{y_0}^{y}dy e^{2F_1{(y)}}   f_3(y) \nonumber\\
    \mathcal{A}_5(y) - \mathcal{A}_5(y_0) =& \int_{y_0}^yd\Bar{y} \Big(2\mathcal{A}_1(\Bar{y})^2 f_3(\Bar{y}) \mathcal{A}_3(\Bar{y}) + \mathcal{A}_1(\Bar{y})^4 f_5(\Bar{y})\Big),
\end{eqnarray}
and $\mathcal{A}_i(0)= \mathcal{A}_i(y_0)$.

These expressions provide us with the eccentricity evolution with respect to the frequency. These can be used for further computations for fluxes and waveforms. In the later sections, we will derive equations for the $f_i$s. Using the derived expression finally, we will find the explicit expressions for the evolving eccentricity.  Once the eccentricity in terms of $y$ is known it can be used to compute shifts in fluxes and waveforms, as well as the orbital dynamics in the inspiral time scale.

\subsection{Aligned and antialigned orbits}

In the last section, we constructed a prescription that can be used to compute eccentricity evolution in terms of reference eccentricity and $y$. Using this we have found the general expression for $e_t$ up to $e_0^5$. Although, we limited ourselves to order $e_0^5$, in principle this can be easily extended further. In this work, we will use 2PN order expressions of $f_1$, $f_3$ and $f_5$.

In such a case we can apply the prescription constructed in the last section, and eccentricity evolution can be found. We find, 


\begin{widetext}
    \begin{equation}
        \begin{split}
           -\frac{6 y}{19} f_1(y) =& 1+\frac{(2833-5516 \nu ) y^2}{3192}+ \frac{1}{456} y^3 (1131 \pi -4 (157 \mu_1 s_1+102 \mu_1 s_2)) \\
           &+\frac{y^4 \left(504 (76191-11662 \nu ) \nu +63504
   \left((178 Q_1+13) s_1^2+165 s_1 s_2\right)-27099209\right)}{9652608} +1 \leftrightarrow 2\\
  \frac{96 y}{145} f_3(y) =& 1 +\left(\frac{124007}{9744}-\frac{62443 \nu }{5220}\right) y^2 + \frac{y^3 (-16011 \mu_1 s_1-16249 \mu_1 s_2 +9375 \pi )}{1740} \\
  &+\frac{y^4 \left(56 \nu  (36048810 \nu
   -10139837)+63504 \left((3816 Q_1+17) s_1^2 + 3799 s_1
   s_2\right)+208352147\right)}{49109760} 1\leftrightarrow 2\\
  -\frac{768 y}{1015} f_5(y) =& 1+\left(\frac{1610201}{56840}-\frac{88789 \nu }{3654}\right) y^2+ \frac{y^3 (-133723 \mu_1 s_1-130031 \mu_1 s_2+82607 \pi )}{6090} \\
  &+\frac{y^4 \left(8 \nu  (3304387534 \nu
   -8209557855)+18144 \left((90884 Q_1+1315) s_1^2 +102889 s_1
   s_2\right)+34118330135\right)}{147329280} + 1\leftrightarrow 2,
   \end{split}
    \end{equation}
\end{widetext}

\begin{figure*}
   \centering
   \includegraphics[width=85mm]{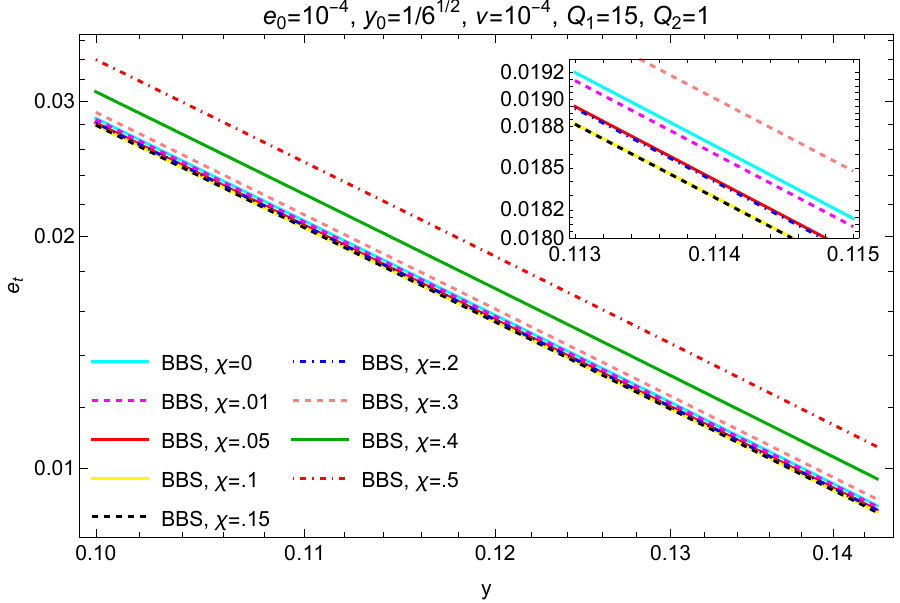}
   \includegraphics[width=85mm]{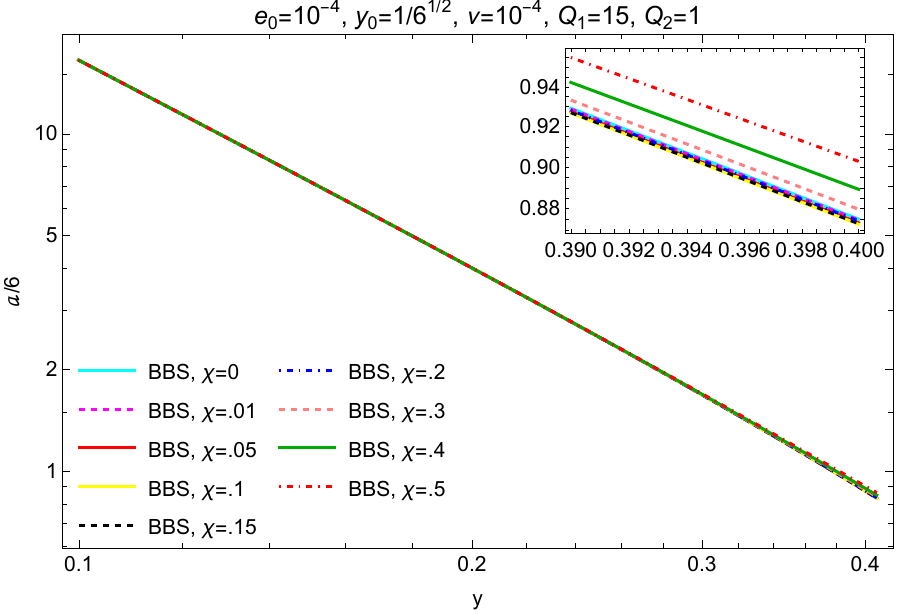}
   \includegraphics[width=85mm]{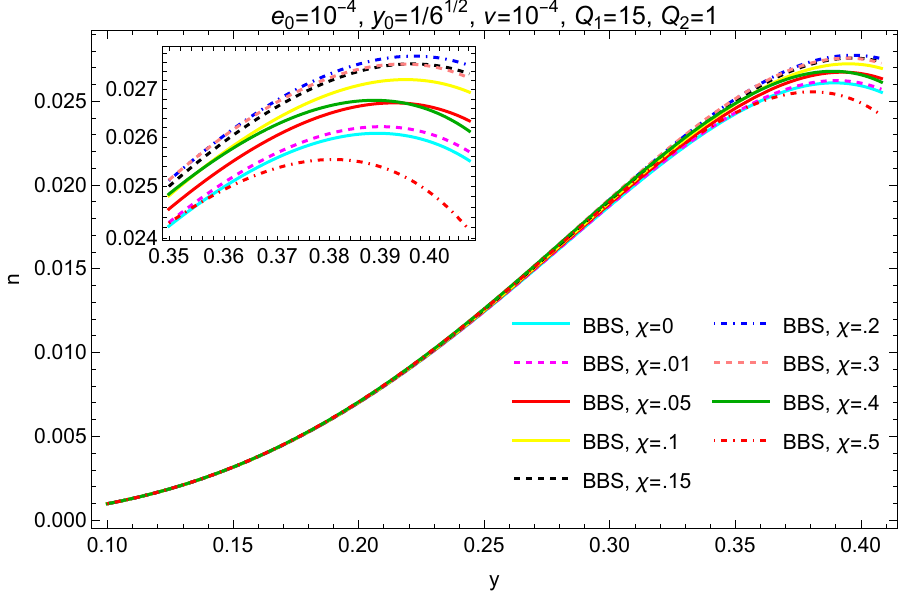}
   \caption{In the above figure we consider an extreme mass ratio inspiral with $\nu=10^{-4}$, with the primary as a boson star of QM $Q=15$, $e_0 =10^{-4}$ at $y_0=1/\sqrt{6}$. In the right figure of the first row we show eccentricity evolution. The eccentricity evolution shows significant EoS dependence. The nonspining binary with same configuration is shown in cyan colour. With decreasing spin the eccentricity values decreases to go below nonspining value before increasing to reach nonspining value. The inset zooms in on the overlapping curves of slowly spinning configuration to demonstrate the QM-spin competition for low spins. In the rest of the panels we show the evolution of $a$ and $n$ defined in Eq. \ref{eq:orbital-quantities}. $a$ and $n$ deviates from non-spinning configuration for larger $Q$ and $\chi$. The deviations are stronger only for larger $y$ as found in other figures also. Clearly EoS dependent deviation is observable in all of the plots.}   
   \label{fig:BBS-spin-effect}
\end{figure*}

where $\chi_i$ and $Q_i$ are the dimensionless spin and the spin induced QM of the $i$th body. With this, we can compute the eccentricity evolution as follows,

\begin{widetext}
\begin{equation}
\begin{split}
   y^{19/6} \mathcal{A}_1(y) =& 1+\frac{(5516 \nu -2833) y^2}{2016}+ \frac{1}{432} y^3 (628 \mu_1 s_1+408 \mu_1 s_2 -1131 \pi ) \\
   &+\frac{y^4 \left(-168 (1015247-613298 \nu ) \nu
   -127008 \left((178 Q_1+13) s_1^2 +165  s_1 s_2\right)+78276085\right)}{24385536} + 1\leftrightarrow 2.
\end{split}
\end{equation}

\begin{equation}
    \begin{split}
       -\frac{608 }{145}y^{19/3} \mathcal{A}_3(y) =&1+\left(\frac{688427}{47502}-\frac{24757 \nu }{2610}\right) y^2  -\frac{19 y^3 (197138 \mu_1 s_1+233322 \mu_1 s_2 -4755 \pi )}{313200} \\
       &+\frac{19 y^4 \left(-336 \nu 
   (5535936 \nu -46054369)+15876 \left((42878 Q_1-1579) s_1^2 +44457 s_1
   s_2\right)-5111769781\right)}{1546957440} +1 \leftrightarrow 2.
    \end{split}
\end{equation}

\begin{equation}
    \begin{split}
    \frac{369664}{59595} y^{38/3}  \mathcal{A}_5(y) =& 1-\frac{19 (1393896140 \nu -2275220379) y^2}{1561865760} - \frac{19 y^3 (2 (72461891 \mu_1 s_1+81507063 \mu_1 s_2)-18507009 \pi )}{124434360}\\
    &+\frac{19 y^4}{17908352804160}
   \Bigg(-1456 \nu  (5007132305 \nu +20844474958)+206388 \Big((59272558 Q_1-1462515) s_1^2 \\
   & +67821313 s_1
   s_2\Big)+114596935508409\Bigg) +1 \leftrightarrow 2.
    \end{split}
\end{equation}

\end{widetext}
where $\mu_i=m_i/(m_1 + m_2)$and $1 \leftrightarrow 2$ represents changing the label of the first and the second body in the binary. In the leading order of $e_0$ for nonspinning orbit this was explored in a few works \cite{Arun:2009mc, Moore:2016qxz}. The current results with the prescription discussed, are consistent with the previous results in the appropriate limits.

\begin{figure*}
   \centering
   \includegraphics[width=85mm]{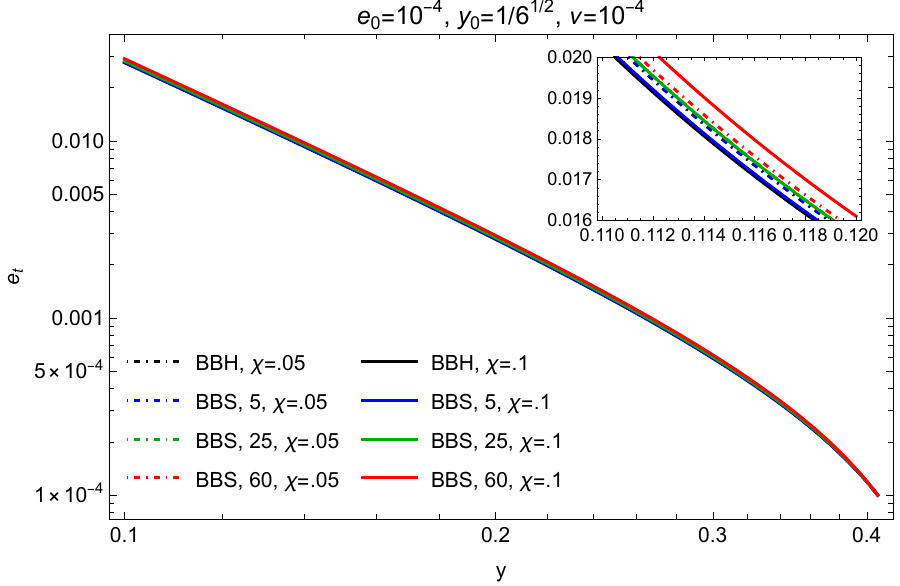}
   \includegraphics[width=85mm]{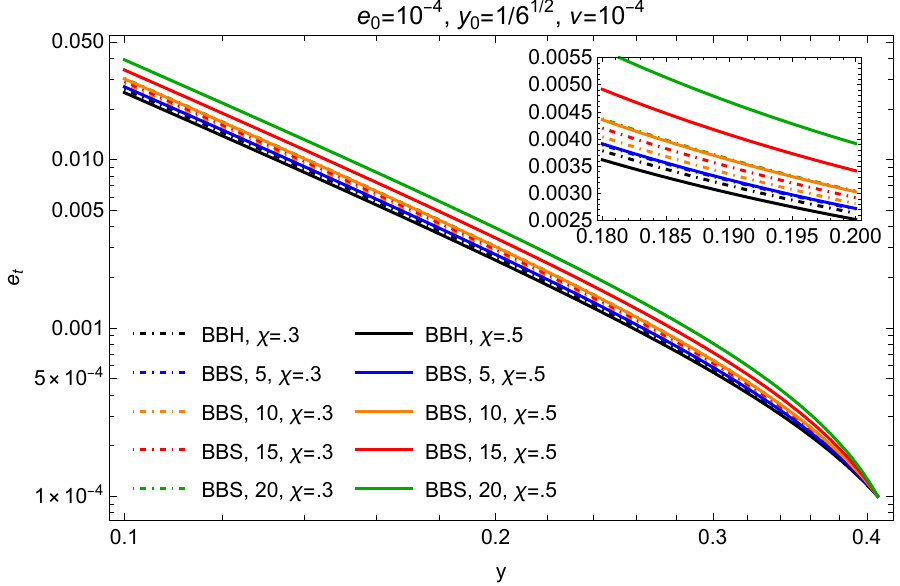}
   \caption{Same as Fig. \ref{fig:BBH-BBS}, but for extreme mass ratio inspirals with $\nu=10^{-4}$. The primary is assumed to be a spinning boson star. Red, green, and blue curves represent $Q=60,\, 25, \, 5$ in the left panel and $Q=20,\, 15, \, 10$ in the right panel. We set the reference eccentricity as $e_0 = e_t(y_0) = 10^{-4}$. In the left panel, we consider primary with dimesionless spin $\chi=.05, .1$ and the right panel, we consider $\chi=.3, .5$.}   
   \label{fig:BBH-BBSEMRI}
\end{figure*}

\section{Impact of quadrupole moment}

With the result found in the last section we study how eccentricities evolve in different systems. In Ref. \cite{Klein:2018ybm} the evolution equations were provided taking non-zero spin induced QM which is also included in the current work. $Q$ is defined such that for a BH $Q=Q_{BH}=1$. In this section we study the impact of $Q$ on the orbital properties due to E-Q coupling. We compare first between binary BBH and binary neutron star (BNS), and then between BBH and binary boson star (BBS). Except for Sec. \ref{sec: large e0}, we will consider small eccentricity at ISCO and evolve it backwards in time and study the impact of QM. This will demonstrate that even for very small observed $e_0$ at ISCO, the impact of EoS of the components has a non-negligible impact on the eccentricity at earlier times. Thereby, demonstrating that the impact of EoS can not be ignored when inferring the initial configurations and formation channels using observed eccentricity.

\subsection{Binary neutron stars}

In Fig.\ref{fig:BBH-BNS} we plot the evolving eccentricity $e_t$ as function of post-Newtonian parameter $y$. For all the systems we set $y_0 = 1/\sqrt{6}$ (representing the ISCO), and evolve the binaries backword in time, i.e. lower $y$ values. 
We set $e_0=e_t(y_0)=10^{-4}$. In the left panel we consider equal mass and equally spinning components, i.e. $m_1=m_2$ and $\chi_1=\chi_2=\chi$. The dashed curve represents dimensionless spin $\chi=.05$ and the dotdashed curves represent $\chi=.1$.
We assume the spin to remain constant through out the inspiral. The values of NS QMs are taken from \cite{Yagi:2013awa}.
From the plots several features can be observed. The evolution depends on the EoS of NSs, very mildly. EoS effect moves the eccentricity value away from the BH values. This spread increases more with increasing spin as can be noticed by comparing the spread of the dashed and the dotdashed curves.
This is because with increasing spin the QM effect becomes larger resulting in larger deviation.

In the right plot of Fig.\ref{fig:BBH-BNS}  we plot the same but for asymmetric mass ratio. The qualitative features stays similar to the equal mass case, except the eccentricity in all the curves is slightly larger compared to the equal mass case.
In the inset we show the $y$ range from $.1$ to $.101$ to zoom into individual curves. The figures demonstrate that the inferred eccentricity from GW observation, in principle, is not sufficient to uniquely determine the binary configuration at formation, even if it is known at which frequency they could have formed. Due to the QM there always will be EoS dependent indeterminacy in the initial configuration. Hence, this will eventually percolate in the population distribution resulting in EoS dependent biases on determining the formation channel. Hence, while studying the population distribution and formational channel this bias should also be folded into the Bayesian analysis to avoid any systematics. However, in BNS as the effect is very mild, this may not impact our current understanding much unless extremely precise population information is present.

In the second row of Fig.\ref{fig:BBH-BNS} we show $n$. For small $y$ the leading order $y$ dependence in Eq. \ref{eq:a-n} is independent of spin and $Q$. As a consequence all the curves overlap. However, for larger $y$ individual curves starts to split from each other due to larger subleading contribution. Since $\chi=.05$ and $.1$ has been considered, with larger $y$, the overlapped curves decomposes into 2 streams only depending mainly on the spin values. For even larger $y$ individual streams decomposes very mildly into different substreams due to $Q$ dependent deviations as can be seen in the subplots. The deviations are primarily present for larger $y$, unlike $e_t$. This is because eccentric orbit is not completely determined by eccentricity alone, it also depends on the latus rectum. Hence, same eccentricity does not lead to exactly same orbital configuration.

\subsection{BNS with subsolarmass NS}

In this section we focus on BNs systems comprising of subsolarmass (SSM) NSs. The detection of SSM compact objects could point to either a new formation channel beyond standard stellar core-collapse scenarios \cite{Metzger:2024ujc} or signal the presence of new physics—such as primordial BHs \cite{Hawking:1971ei, Carr:1974nx, Carr:1975qj}. Such a discovery would carry profound implications for astrophysics, cosmology, and fundamental physics.
Detecting SSM objects in compact binary mergers is thus a key objective for current and upcoming GW observatories \cite{Branchesi:2023mws}. Although several GW searches using LVK data have targeted coalescing binaries with at least one SSM component, no definitive detections have been reported \cite{LIGOScientific:2018glc, LIGOScientific:2019kan, Nitz:2021mzz, LIGOScientific:2021job}. Notably, candidate SSM binary events were identified \cite{LIGOScientific:2021job, Morras:2023jvb}, though were not confirmed due to high false alarm rates.

\begin{figure*}
   \centering
   \includegraphics[width=85mm]{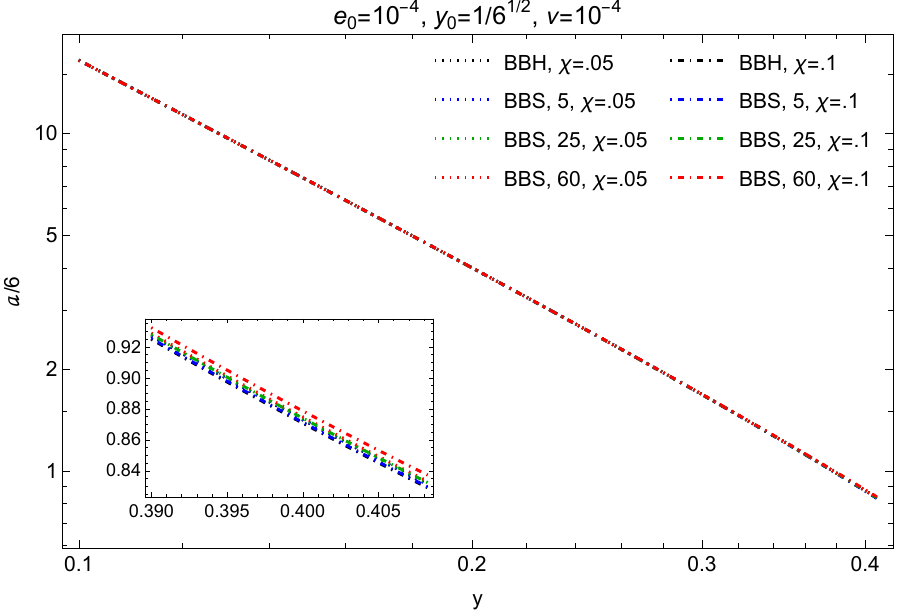}
   \includegraphics[width=85mm]{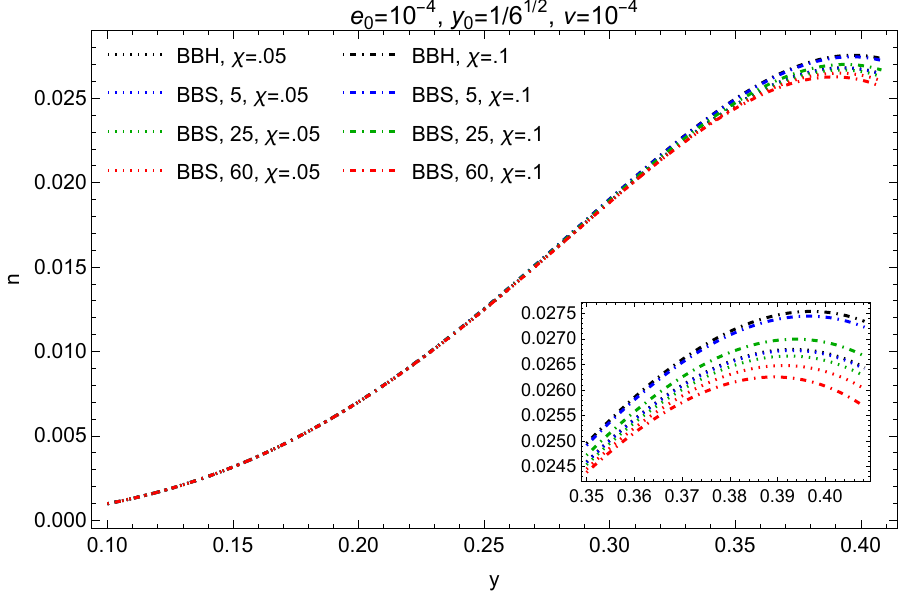}
   \includegraphics[width=85mm]{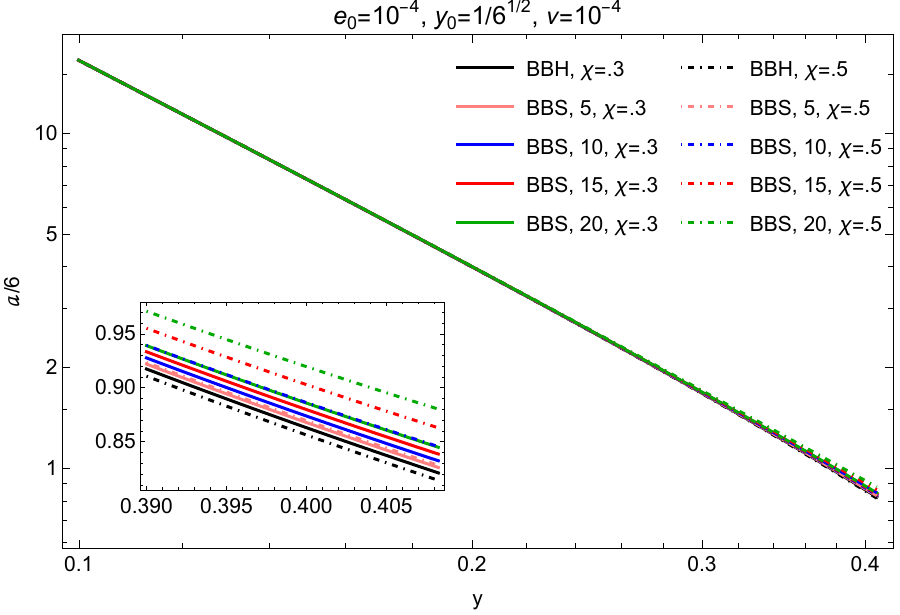}
   \includegraphics[width=85mm]{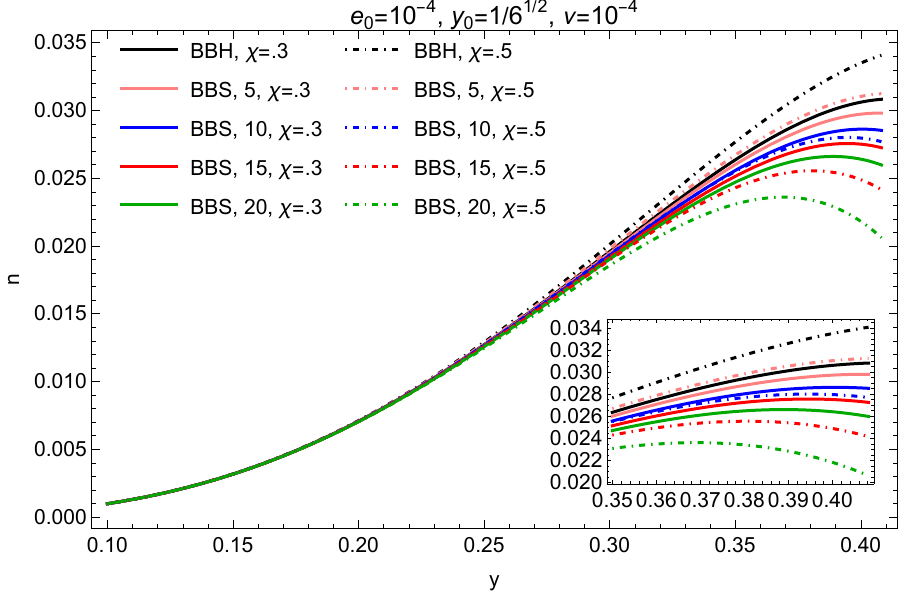}
   \caption{In the above figure we show the evolution of $a$ and $n$ defined in Eq. \ref{eq:orbital-quantities}. The configurations are same as in Fig. \ref{fig:BBH-BBSEMRI}. $a$ and $n$ deviates from BH values for larger $Q$ and $\chi$. The deviations are stronger only for larger $y$.}   
   \label{fig:BBH-BBSEMRI-a-n}
\end{figure*}

In this context Ref. \cite{Crescimbeni:2024qrq} argued, data from the O4–O5 observing runs could enable tidal deformability measurements precise enough to confirm or rule out the existence of light NSs composed of strange quark matter \cite{Weber:2004kj, Haensel:1986qb}.
Such a result would significantly advance our understanding of the nuclear EoS at ultra-high densities \cite{Prakash:1995uw, Lattimer:2000nx, Agathos:2015uaa}. As noted in prior work, this opportunity arises because quark stars in the SSM range are expected to have significantly smaller radii—and therefore much smaller tidal deformabilities—than standard NSs. Conversely, the overall tidal deformability of SSM NSs is large \cite{Crescimbeni:2024cwh}, making it more accessible to measurement than that of heavier NSs. Recently Ref. \cite{Doroshenko:2022nwp} has claimed to have observed subsolar mass NS also. Existences of such objects can impact our understanding of formational channels significantly.

In Ref. \cite{Sham:2014kea} relations between tidal deformability and QM for low-mass NSs and quark stars were investigated. The relations for low-mass NSs become increasingly sensitive to the EoS as the dimensionless tidal deformability grows, showing significant deviation from the incompressible limit. Similar to the argument in Ref. \cite{Crescimbeni:2024qrq} about observability of SSM via tidal Love number, searching for them using QM also arises for slowly rotating NSs. As a consequence the E-Q coupling could open up more possibilties to confirm/exclude the existence of SSMs.

Keeping these aspects in mind we explore the E-Q coupling for SSM objects. In Fig. \ref{fig:BBH-BNS-subsolar} we show the evolution of $e_t$ and $n$ for equal mass BNSs comprising of $.3M_{\odot}$, $.5M_{\odot}$, and $.7M_{\odot}$. The value of $Q$ has been taken from Ref. \cite{Yagi:2013awa}. 
$n$ is sensitive to EoS for larger $y$ whereas $e_t$ is sensitive to it for smaller $y$, for a fixed $e_0$ at $y_0=1/\sqrt{6}$. This implies that eccentricity evolution of SSM NSs are more sensitive to EoSs compared to massive NSs. They all show deviations from BBH binaries. EoSs that allow larger $Q$ induces larger deviation.

\subsection{Binary with equally massive Boson stars}

Similar to the Fig. \ref{fig:BBH-BNS}, in Fig. \ref{fig:BBH-BBS} we show the eccentricity evolution of BBH and BBS containing similarly massive and similarly spinning components. The QMs for boson stars were computed in \cite{Ryan:1996nk, Vaglio:2022flq}. Unlike BHs where the QM depends only on the mass and spin, for boson stars along with mass and spin they also depend on the properties of the bosonic fields, i.e. EoS. In \cite{Ryan:1996nk, Vaglio:2022flq} the dependence on EOS comes through different values of effective mass parameter that is a combination of different field parameters.
Qualitatively the features of Fig. \ref{fig:BBH-BBS} is similar to the Fig. \ref{fig:BBH-BNS}. However, in BBS cases the deviations from the BBH curves are more significant compared to the BNS. This happens solely due to the large values of the boson star QMs.
This opens up the possibility to test the ``exotic natures" of the compact objects in binary from eccentricity evolution. Since the eccentricity evolution has EoS dependence, it can be used to rule out the extreme EoSs. This can be done by taking observational constraint on $e_0$ from GW, and integrating it backwards with different EoSs, i.e. different QMs. These different EoSs will result in different intial orbital cinfiguration ``at formation". From the astrophysically motivated formation chanels some of these configurations will fall into the category of ``unphysical" initial orbital configuration. Such intial configurations and as a result the corresponding EoSs can be ruled out. Hence, GW observations along with population studies has the potential to jointly limit how much the QMs can differ from that of a BH. This can independently provide us with probes to test the BHness of an observed binary.

In Fig. \ref{fig:BBH-BBS-a-n} we show the evolution of $a$ and $n$ defined in Eq. \ref{eq:orbital-quantities} for similar configurations. $a$ and $n$ deviates from BH values for larger $Q$ and $\chi$. This explains why in BNS this effect is less significant. The deviations in $n$ are primarily present for larger $y$, unlike $e_t$ as was also the case in Fig.\ref{fig:BBH-BNS}. This is because eccentric orbit is determined not only by eccentricity but also depends on the latus rectum. Hence, same eccentricity can lead to different orbital configuration if latus rectum is different. This can also be understood from Eq. \ref{eq:a-n} and Eq. \ref{eq:gamma}. For larger $y$, $\gamma$ that has $Q$ dependence contributes to $a$ and $n$. However, for smaller $y$ they are dominated by the leading order contribution that is independent of $Q$ and $\chi$.

\subsection{EMRI with Boson star primary}

In the left panel of the first row of Fig. \ref{fig:BBS-spin-effect} we consider an extreme mass ratio inspiral (EMRI) with $\nu=10^{-4}$, with the primary as a boson star with QM $Q=15$. With a fixed $Q$ and fixed $e_0 =10^{-4}$ at $y_0=1/\sqrt{6}$ we demonstrate how the spin effect on eccentricity evolution changes. For comparison we also show the nonspining binary with same configuration in cyan colour. Note, with decreasing spin the eccentricity values decreases and even goes below nonspining value and then increases to reach nonspining value for very small spins. This is because a large $Q$ introduces a competition between $1.5$-PN spin effect, which does not have $Q$ contribution, and $2$-PN spin effect comprising of $Q$. Hence for a system with multipolar structure deviating strongly from BH values can have interesting dynamics that is different from that of a BH. The curves of $a$ and $n$ in the other panels show similar behavior like in the Fig. \ref{fig:BBH-BBS-a-n}. However, due to the largeness of $Q$ for BBSs, the deviations are much larger.

In Fig. \ref{fig:BBH-BBSEMRI}, similar to Fig. \ref{fig:BBS-spin-effect}, we venture into the EMRI where the primary component is either a BH or a boson star. We take the mass ratio $\nu=10^{-4}$. In the left panel we restrict ourselves in the slowly spinning limit. Due to the largeness of the $Q_i$, here we find the deviations from EMRIs with BH primary can be comparatively larger from BNS. In the right panel where we consider larger spins, we find similar qualitative feature. However the deviations are much larger. Hence, in EMRIs comprising with supermassive objects (observable with LISA) QM can induce observable impact even from eccentricity evolution.
In several works it has been demonstrated that QMs can modify the fluxes, resulting in dephased waveforms. Under circular orbit assumptions the observability has also been studied \cite{Krishnendu:2017shb,Datta:2019euh}. Here for the first time we show that due to the E-Q coupling new QM dependent features can arise in the waveforms. It remains to be seen how the observability of QM changes in the presence of the eccentricity.

In Fig. \ref{fig:BBH-BBSEMRI-a-n} we show $a$ and $n$ for similar configuration that of in Fig. \ref{fig:BBH-BBSEMRI}. In the first row which focuses in low spins show less significant deviation from BBH. However, in the second row, the large spin introduces larger deviation. This is consistent with our previous figures as well as with the governing equations.
In the EMRIs with supermassive objects, it is expected that the spin of the primary can be large. In our work we assumed $e_0 = 10^{-4}$, which is also conservative. It is expected that the EMRI binaries could have significantly larger residual eccentricities \cite{LISA:2022yao}. In such cases the deviations can be much larger. However, we restrict ourselves to very conservative values, otherwise the small eccentricity expansion can fall short of accurately describing the real scenario. The E-Q interaction and its impact on the evolution is a completely new feature, that requires further investigations. 

In binary systems involving boson stars, the orbital dynamics are influenced by a range of effects beyond the leading-order quadrupole moment effect. Additional contributions can significantly affect the binary evolution. Notably, scalar-wave emission provides an extra channel for energy and angular momentum dissipation, thereby accelerating or modifying the inspiral relative to compact-object binaries. Resonant phenomena may also arise when the orbital frequency coincides with intrinsic oscillation modes of the boson star, resulting in enhanced energy transfer between the orbit and the stellar configuration. Moreover, tidal interactions, stemming from the extended and self-gravitating structure of boson stars, can imprint further corrections to the orbital motion. Collectively, these effects underscore the rich phenomenology of boson star binaries that is also sensitive to EoS of it that goes beyond QM correction.

\subsection{EMRI with Neutron star as primary}

\begin{figure}
   \centering
   \includegraphics[width=85mm]{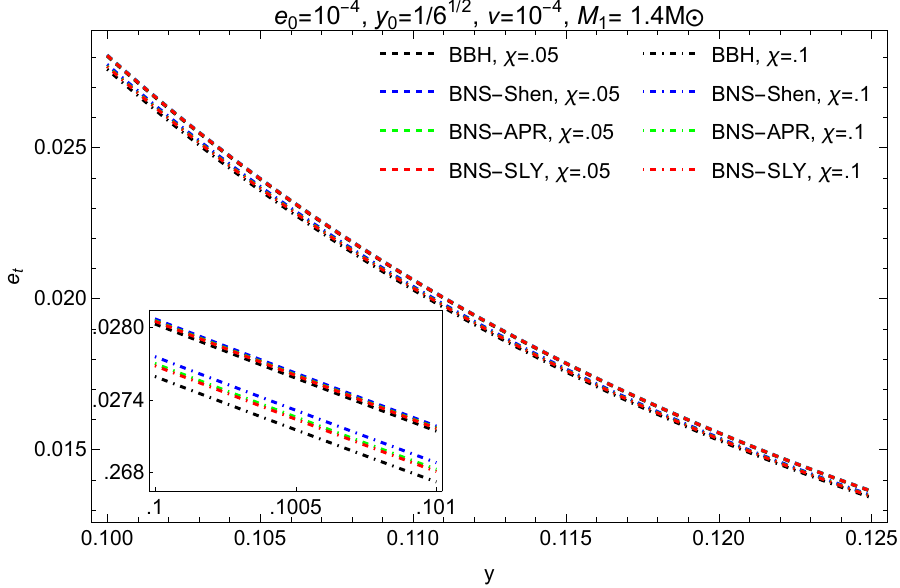}
   \includegraphics[width=85mm]{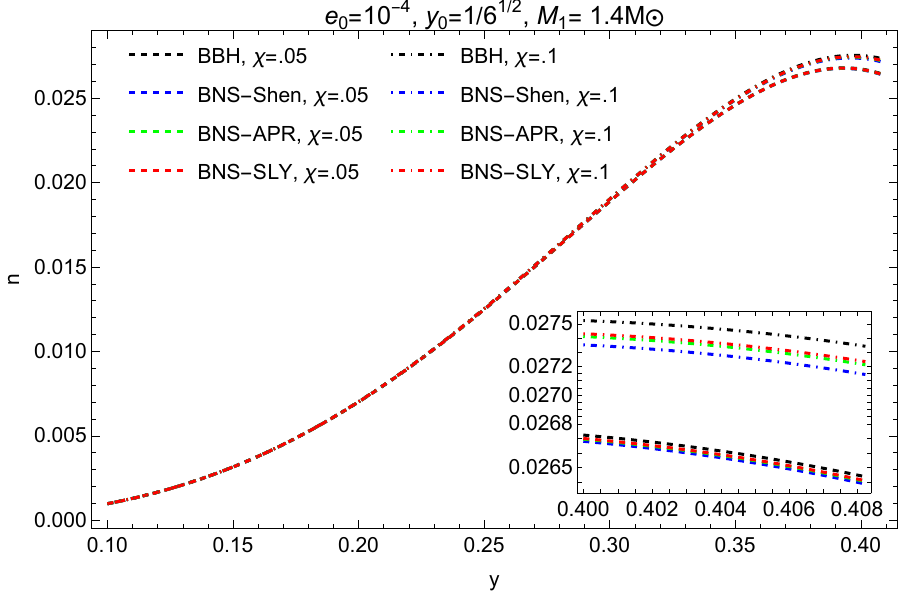}
   \caption{We show the evolution for an EMRI with NS as primary and the secondary as subsolar mass primordial BH. The mass ratio considered is $\nu=10^{-5}$. Both bodies are equally spinning with dimensionless spin $\chi$.}   
   \label{fig:BBH-BNSEMRI}
\end{figure}

In the left panel of Fig. \ref{fig:BBH-BNSEMRI} we construct EMRIs where a neutron star is the primary. Hence, these binaries will be of total mass of the order of $\sim M_{\odot}$. These binaries can form if a subsolarmass primordial BH gets captured by a NS. These systems will then form EMRIs in LIGO-Virgo band. Due to the mass ratio, the eccentricity growth for all configurations is different from the ones in Fig. \ref{fig:BBH-BNS}. However, due to the smaller values of QMs the deviation from BBH value is mild unlike the EMRIs with boson stars in Fig. \ref{fig:BBH-BBSEMRI}. In the right panel we show  $n$. Here also we see that with larger $y$ first the overlapped curves branch into two streams due to the spin difference and for even larger $y$ they split into different branches according to the EoSs. EoS dependence in the evolution of $a$ was found to be less significant.

\subsection{Large eccentricity limit }
\label{sec: large e0}

The analytic expressions derived, while reliable for binaries with small eccentricities, are not sufficient to capture the dynamics of systems with large eccentricities. To assess the influence of the QM in the high-eccentricity regime, we numerically integrate the evolution equation $e_t'(y) \;=\; \frac{\dot{e}_t}{\dot{y}}$ while fixing the initial orbital parameter to $y_0 = 1/\sqrt{6}$. In all cases, we consider a set of reference eccentricity values, namely $e_0 = 0.001,\, 0.01,\, 0.1,$ and $0.5$. 

\begin{figure*}
   \centering
   \includegraphics[width=85mm]{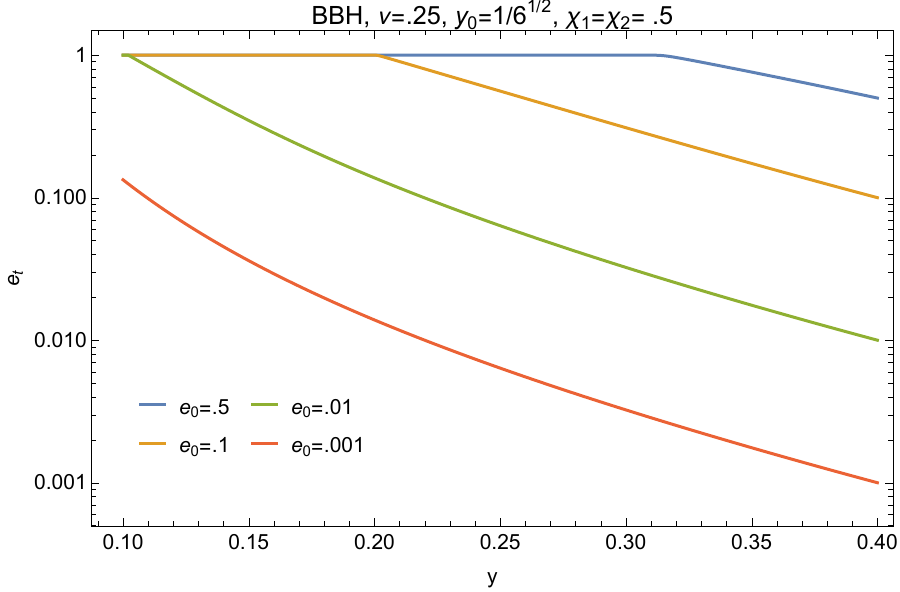}
   \includegraphics[width=85mm]{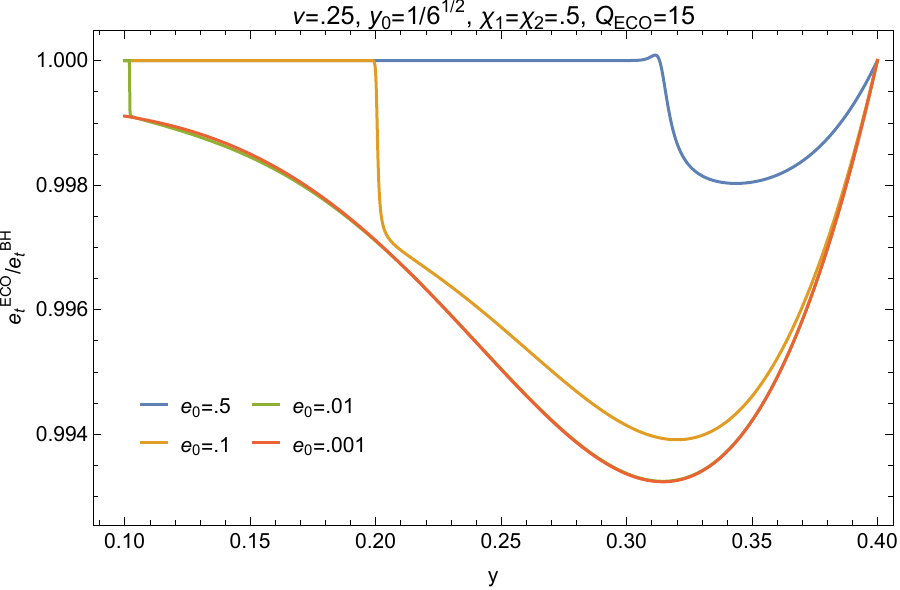}
   \includegraphics[width=85mm]{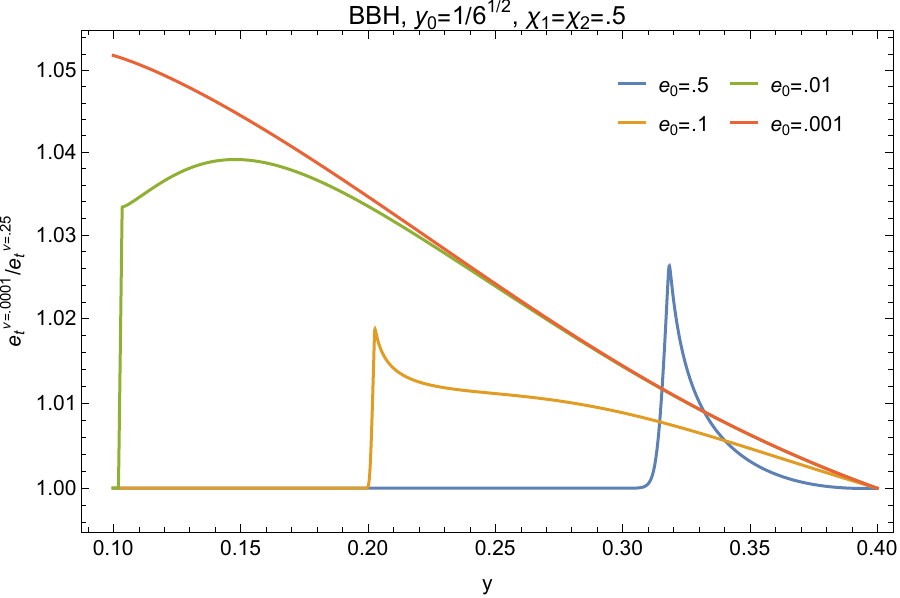}
   \includegraphics[width=85mm]{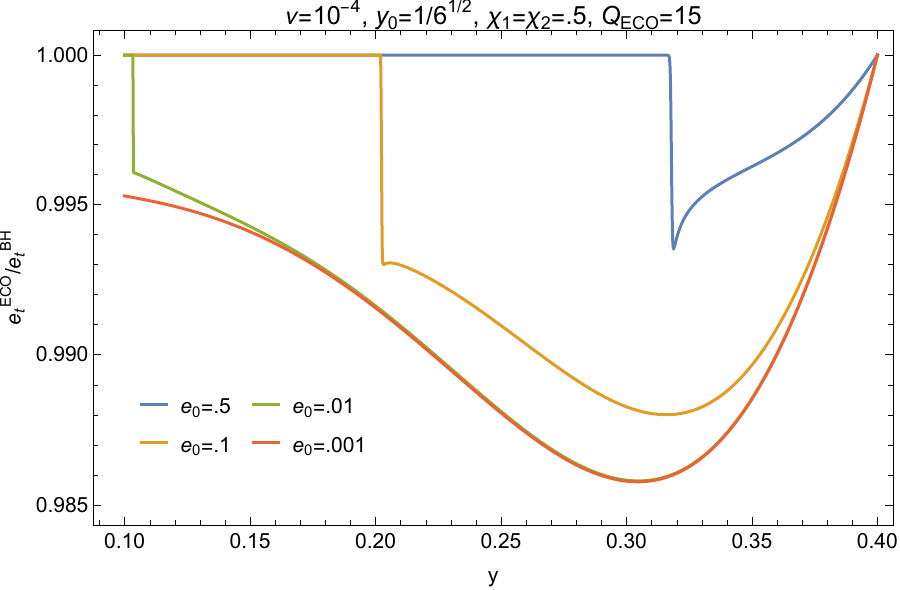}
   \caption{In the above figure we compare eccentricity evolution of different binary composition for larger eccentricity system, by numerically integrating the evolution equations. In each panel, considered $e_0= .5,\, .1, \, .01, \, .001$. Spin values in each panels are $\chi_i =.5$. In the upper left panel the evolution for equal mass BBH is shown. In the upper right panel evolution of binary ECO $(e_t^{\rm ECO})$ with $Q_i=15$ is compared with upper left panel $(e_t^{\rm BH})$. The bottom left panel compares BBH EMRI $(e_t^{\nu= .0001})$ with upper left panel $(e_t^{\nu= .25})$ and the right bottom panel compares EMRI ECO with $Q_i=15$ with EMRI BBH in the bottom left panel.}   
   \label{fig:large-e0}
\end{figure*}

The results are summarized in Fig. \ref{fig:large-e0}. In the upper left panel, we present the backward eccentricity evolution for an equal-mass BBH configuration with both component spins fixed to $0.5$. As expected, systems with larger initial eccentricities $e_0$ rapidly evolve towards the extremal value $e_t = 1$. In the upper right panel, we compare the $e_t$ evolution of the BBH case with that of an equal-mass ECO binary characterized by $Q_i = 15$. The bottom left panel illustrates a comparison between an equal-mass BBH and a BBH EMRI configuration, while the bottom right panel compares a BBH EMRI against an ECO EMRI with $Q_i = 15$. In each case, the change observed in the curves corresponds to the moment when the eccentricity parameter $e_t$ of one of the binaries reaches unity.

\section{Discussion}

The first result of the paper is the formulation of the prescription that can be used to compute the eccentricity evolution of a spinning binary to very high order and studying the impact of EoS on the evolution. It provides us with a frequency-dependent evolution of eccentricity in terms of a reference eccentricity. We also discussed how it can be extended to higher orders iteratively. We considered the spins to be (anti)aligned to simplify the calculations. With this we provide 2PN results upto $\mathcal{O}(e_0^5)$. Hence, these results can be used to model GW waveforms considering eccentricity-spin coupling.

We also consider nonzero spin induced QM. We used this opportunity to study the effect of the binary component properties on the eccentric orbit through the spin induced QM. We demonstrated for the first time, that the eccentricity evolution depends on the Equation of the state of NSs, though very mildly, in a BNS. We find that the EoS effect deviates the eccentricity values from the BBH values. With the increasing spin the deviation increases. This happens because the QM effect proportional to spin square.
We found that the features stay the same qualitatively for asymmetric binaries also.
We find similar features when binary boson stars are considered. The deviation in this case are larger since the QMs of boson stars can be larger.
The observed deviations for both the binary boson stars and NSs demonstarte that with the inferred eccentricity from GW observation it is not possible to uniquely determine the binary configuration at earlier times, unless the EoS of the compact object is known to ``sufficient accuracy". Due to the QM the EoS will always introduce indeterminacy in the initial configuration. We argue that this may eventually percolate in the population distribution, resulting in EoS dependent biases on determining the formation channel. Therefore, to avoid any systematic biases, this EoS dependent biases should also be folded into the Bayesian analysis while studying the population distribution and formational channel. 

We also argue that the E-Q coupling can therefore open up another possibility to test the ``exotic natures" of the compact objects in binary from eccentricity evolution itself. From observational constraint on $e_0$ found from GW measurement and correlating it with astrophysically motivated formation channels jointly constraints can be put on how much the QMs can differ from that of a BH. This possibly will open up another prospect to measure the BHness of a compact object. This requires further investigations. Except for EMRIs with BS as primary we find effect of QM on $a$ to be less significant.

\section*{Acknowledgements}

We thank Khun Sang Phukon for suggesting to look into the effect of QM on binary evolution. We want to thank K. G. Arun, Prasanta Char, and Antoine Klein for the very useful discussions. We also like to thank Chandra Kant Mishra, and Khun Sang Phukon for reading an earlier draft and providing us with valuable comments.

\appendix

\section{Higher order terms of $e_0$}
\label{app:Higher order e0}

From the formalism, it is straightforward to check that the term in the expression of $e_t$ corresponding to $e_0^n$ will consist of a term, 

\begin{equation}
    \begin{split}
        \mathcal{O}(e_0^n) =& \frac{\mathcal{A}_1(y)  }{\mathcal{A}_1(y_0)^n}  e_0^n \left[ \mathcal{A}_n(y) -\mathcal{A}_n(y_0)\right]
      + {\rm other\,\, terms} \\
      \left[ \mathcal{A}_n(y) -\mathcal{A}_n(y_0)\right] \sim& \int_{y_0}^y \mathcal{A}_1(\Bar{y})^{n-1}  f_{n}(\Bar{y}) d\Bar{y} .
    \end{split}  
\end{equation}

Although it is time-consuming to find the complete expression for $\mathcal{O}(e_0^n)$, it is easier to evaluate the above integral to find an approximate result. For exactness rest of the terms must be found in the manner discussed previously. In this way, we will find the approximate result for several higher-order terms.

\begin{widetext}
    \begin{equation}
    \begin{split}
       \frac{6144 y}{7105} f_7 =& 1+\frac{(316165851-264138140 \nu ) y^2}{7161840}+ \frac{y^3 (5471123 \pi -3 (2982097 \mu_1 s_1+2868059 \mu_1 s_2 ))}{255780} \\
       &+\frac{y^4}{3609567360} \Bigg(168 \nu 
   (10167524710 \nu -23951330779)+31752 \left((2165576 Q_1+32185) s_1^2 +2319871 s_1
   s_2\right) \\
   & +2437574034215\Bigg)  + 1 \leftrightarrow 2
    \end{split}
\end{equation}

\begin{equation}
    \begin{split}
       \frac{-49152 y}{49735} f_9 =& 1+\left(\frac{5964533}{99470}-\frac{210863 \nu }{4263}\right) y^2+ \frac{y^3 (523442279 \pi -2520 (341006 \mu_1 s_1+326307 \mu_1 s_2 ))}{17904600} \\
       &+\frac{y^4}{8422323840} \Bigg(56 \nu 
   (139190804970 \nu -355549443973)+1778112 \left((127449 Q_1+1915) s_1^2 +133304 s_1
   s_2\right) \\
   &+11805696909955\Bigg) +1 \leftrightarrow 2
    \end{split}
\end{equation}

\begin{equation}
    \begin{split}
     \frac{-116736 y^{19}}{7105}\mathcal{A}_7 \sim & 1 -\frac{19 (18320575 \nu -31972557) y^2}{15218910}-\frac{19 y^3 (2 (6715321 \mu_1 s_1+7154757 \mu_1 s_2)-2906491 \pi )}{8184960} \\
     & +\frac{19 y^4}{54143510400} \Bigg(672
   \nu  (29028405 \nu -1269716549)+15876 \left((3066462 Q_1-27995) s_1^2+ 3467417 s_1
   s_2\right) \\
   &+1270466088403\Bigg) +1 \leftrightarrow 2
    \end{split}
\end{equation}

\begin{equation}
    \begin{split}
      \frac{1245184 y^{76/3}}{49735}  \mathcal{A}_9 \sim & 1 -\frac{19 (49371280 \nu -87233129) y^2}{31333050} -\frac{19 y^3 (280 (6976192 \mu_1 s_1+7360869 \mu_1 s_2 )-445321137 \pi )}{899706150} \\
      & +\frac{19 y^4}{606407316480}
   \Bigg(840 \nu  (3889118870 \nu -34594102133)+222264 \left((3323474 Q_1-23425) s_1^2+ 3626619 s_1
   s_2\right) \\
   &+30645477963727\Bigg) +1 \leftrightarrow 2
    \end{split}
\end{equation}

\end{widetext}

\bibliography{references}

\end{document}